\documentclass[aps,prd,amssymb,cite,
amsfonts,epsf,preprintnumbers,nofootinbib,superscriptaddress]{revtex4}

\usepackage[dvips]{graphicx}
\usepackage{bm,latexsym,amsmath,amssymb,amsfonts}
\usepackage[usenames,dvipsnames]{color}
\usepackage[colorlinks=true,linkcolor=blue]{hyperref}
\usepackage{color}
\usepackage{ulem}
\usepackage{epsfig}
\usepackage{braket}
\usepackage[mathscr]{eucal}
\usepackage{cancel}
\usepackage{mathrsfs}
\usepackage{pgf,tikz}
\usepackage{slashed}
\usetikzlibrary{arrows,automata}

\definecolor{mypink1}{rgb}{0.858, 0.188, 0.478}
\definecolor{mypink2}{RGB}{219, 48, 122}
\definecolor{mypink3}{cmyk}{0, 0.7808, 0.4429, 0.1412}
\definecolor{mygray}{gray}{0.6}
\definecolor{pptbg}{rgb}{0.961,0.945,0.863}

\newcommand{\be}[1]{\begin{equation} \label{#1}}
\newcommand{\ee}{\end{equation}}
\newcommand{\bea}{\begin{eqnarray}}
\newcommand{\eea}{\end{eqnarray}}
\newcommand{\nn}{\nonumber}

\newcommand{\cT}{\mathcal{T}}
\newcommand{\hreff}[1]{\href{#1}{\color{blue}{#1}} }
\newcommand{\sech}{\,{\rm sech}\,}

\begin{document}

\title{Temperature of a steady system around a black hole }

\author{Hyeong-Chan Kim}
\affiliation{School of Liberal Arts and Sciences, Korea National University of Transportation, Chungju 380-702, Korea}

\email{hckim@ut.ac.kr}



\begin{abstract}
We study the issue of temperature in a steady system around a black hole event horizon, contrasting it with the appearance of divergence in a thermal equilibrium system. 
We focus on a spherically symmetric system governed by general relativity, particularly examining the steady state with radial heat conduction.
Employing an appropriate approximation, we derive exact solutions that illuminate the behaviors of number density, local temperature, and heat in the proximity of a black hole.
We demonstrate that a carefully regulated heat inflow can maintain finite local temperatures at the black hole event horizon, even without considering the back-reaction of matter.
This discovery challenges conventional expectations that the local temperature near the event horizon diverges in scenarios of thermal equilibrium. 
This implications shows that there's an intricate connection between heat and gravity in the realm of black hole thermodynamics.
\end{abstract}

\maketitle
 
\section{Introduction}
Consider an observer who watches a thermodynamic system interacting with a static black hole described by spatially varying metric components, \(g_{ab}\), in Einstein gravity. If he/she measures the temperature of the system in thermal equilibrium, it ticks the local Tolman temperature~\cite{Tolman,Tolman2},
\be{Tolman1}
\Theta(x^i) = \frac{T_\infty}{\sqrt{-g_{00}(x^i)}},
\ee
where \(g_{00}\) and \(T_\infty\) represent the time-time component of the metric on the static geometry and the physical temperature at the zero gravitational potential hypersurface usually located at spatial infinity, respectively.
There were arguments~\cite{Lima:2019brf,Kim:2021kou} for the modification of the original form of the temperature, however, the local Tolman temperature is generally accepted because of the universality of gravity~\cite{Santiago:2018kds,Santiago:2018lcy} and the maximum entropy principle~\cite{Sorkin:1981,Gao:2016trd}. 
The generalization of the formula to systems in stationary spacetimes was also sought~\cite{Buchdahl:49}.

However, a potential issue arises when applying this formula to an observer passing through the event horizon. 
As the observer approaches the event horizon, 
$g_{00}$ tends to zero, leading to an infinite local temperature according to the Tolman temperature formula~\eqref{Tolman1}. 
This seems problematic because a freely falling observer, according to the principle of equivalence in general relativity, should not notice the presence of an event horizon.
The discussion suggests a need to investigate how the divergence of local temperature at the horizon occurs, especially when the system is not in thermal equilibrium. 
We aim to explore the temperature of a steady thermal system where heat flows around the black hole, providing a more nuanced understanding of temperature measurements near the event horizon.

\vspace{.1cm}
Ever since Hawking deduced the thermodynamics of black holes~\cite{Hawking:1974sw}, the close connection between gravity and thermodynamics has been investigated in a diverse era of theoretical physics~\cite{Bekenstein:1974ax,Jacobson:1995ab,Padmanabhan:2009vy,Verlinde:2010hp,Carlip:2014pma,cocke,Kim:2019ygw,Lee:2008vn,Lee:2010bg}.
Examining self-gravitating systems in thermal equilibrium has been one of those efforts for decades, aiding our understanding of astrophysical systems. Especially, the entropy of a spherically symmetric self-gravitating radiation and its stability were calculated in a series of researches~\cite{Sorkin:1981,Gao:2016trd,Roupas:2014nea,Kim:2019ygw}. Those studies have shown that requiring the maximum entropy of self-gravitating radiation in a spherical box reproduces the Tolman-Oppenheimer-Volkoff equation for hydrostatic equilibrium~\cite{Oppenheimer,Tolman3,Cocke}.
Going beyond thermal equilibrium for these fields is anticipated, which may require a deep study of non-equilibrium relativistic thermodynamics.

Historically, relativistic thermodynamics has been pursued along two tracks. 
First is the Israel-Stewart theory~\cite{IS1,IS2,IS3}, which directly generalizes Eckart's thermodynamics~\cite{Eckart:1940aa} for irreversible processes to be compatible with general relativity. Second is Taub~\cite{Taub54} and Carter's axiomatic approach~\cite{Carter72,Carter73,Carter89}, which begins with a Lagrangian-like function, \(\Lambda\). 
Both approaches are known to have the same degrees of generality and are equivalent in the limit of linearized perturbations about a thermal equilibrium state.
As noticed in Ref.~\cite{Priou1991,Gavassino:2021kpi}, Carter's theory of relativistic thermodynamics and the Israel and Stewart formalism must be integrated into a more comprehensive theory of thermodynamics. 
The Israel and Stewart formalism has also been demonstrated to be stable and causal~\cite{Hiscock1983}. 
The formulation was further developed to include dissipations and particle creations in the formalism~\cite{Andersson:2013jga,AnderssonNew}. 
Typically, relativistic heat conduction theory was believed to be incomplete~\cite{Andersson2011}. 
Only recently, the binormal equilibrium condition was proposed to compensate the incompleteness~\cite{LK2022}.
The steady thermal state were studied in Refs.~\cite{Kim:2023lta} based on the result.

We survey the heat conduction equation, which is usually called the relativistic analogy of the Cattaneo equation, based on the action formalism for thermodynamics for two fluids. The variational formulation of relativistic thermodynamics stems from the assumption that the matter flux \(n^a\) and the entropy flux \(s^a\) are two independent fluids interacting with each other. The particle number in the system is assumed to be large enough that the fluid approximation is applied and there is a well-defined matter current \(n^a\).
A typical system of this kind is laboratory superfluids~\cite{Carter94,Andersson11}. For a condensed review of this subject, consult Andersson and Comer~\cite{AnderssonNew}.

In this model, the entropy flux \(s^a\) is, in general, not aligned with the particle flux \(n^a\). 
The misalignment associated with the heat flux \(q^a\) leads to entropy creation. 
The formulation is described in Eckart decomposition where the observer's four-velocity \(u^a\) is parallel to the number flux. Explicitly, given the number density \(n\), the entropy density \(s\), and the heat flux \(q^a\), the particle number and the entropy fluxes are
\be{na}
n^a \equiv n u^a, \qquad s^a \equiv s u^a + \varsigma^a; \qquad \varsigma^a \equiv \frac{q^a}{\Theta},
\ee
where \(q^a u_a = 0\). 
With this construction, the heat flux \(q^a\) denotes the deviation of the entropy flux relative to the number flux. 
This procedure defines the heat uniquely irrespective of the choice of coordinate system, at least for this two-fluid model. 
Note also that, to this comoving observer, the heat appears as an off-diagonal element of the stress tensor $T_{ab}$:
\be{heat:comoving}
q^c = -u^a(g^{bc}+ u^bu^c) T_{ab} .
\ee
Therefore, the heat flux \(q^a\) is the energy flux measured by a comoving observer with the matter. 
Starting from the master function \(\Lambda(n,s,\varsigma)\), the energy density can be obtained by the Legendre transformation, \(\rho =u^au^bT_{ab}= \varsigma \vartheta - \Lambda\). The variational law, now, presents the first law of thermodynamics:
\be{1st law}
d\rho(n,s,\vartheta) = \chi dn + \Theta ds + \varsigma d\vartheta, \qquad \vartheta \equiv \beta q,
\ee
where \(\varsigma(n,s,q)\) and \(\vartheta(n,s,q)\) are a pair of thermodynamic quantities that represent the deviation from thermal equilibrium, and \(\chi\) denotes the chemical potential.

One of the main results for relativistic thermodynamics is the study of thermal equilibrium. Even when the geometry is dynamical, thermal equilibrium with its neighborhoods is characterized by the vanishing of the Tolman vector \(\cT_a\),
\be{Tolman}
\cT_a=0; \qquad \cT_a = \frac{d}{d\tau} (\Theta u_a) + \nabla_a \Theta,
\ee
where \(d/d\tau \equiv u^b \nabla_b\) and \(\nabla_a\) denotes the covariant derivative for a given geometry. The Tolman temperature~\eqref{Tolman1} appears when the geometry is static additionally. For a two-fluid system with one number flow \(n^a\), the other equation characterizing thermal equilibrium is Klein's relation~\cite{Klein49}\footnote{As noted in Ref.~\cite{Kim:2021kou}, Klein's relation may not hold for models with more than three fluids.} \(\gamma^{ab} \mathcal{K}_b = 0\), where \(\mathcal{K}_b \equiv d(\chi u_b)/d\tau +\nabla_b \chi\).
The stability and causality of the thermal equilibrium state were also analyzed~\cite{Hiscock1987,Olson1990,LK2022}. 

\vspace{.1cm}
In traditional thermodynamics, heat is closely linked to temperature difference. 
Heat flows between two neighboring systems A and B if and only if there is a temperature difference. 
If heat does not flow between the two, they are in equilibrium.
In the presence of gravity, thermal equilibrium is characterized by the Tolman temperature gradient~\eqref{Tolman}. 
Therefore, Tolman's relation should hold between A and B even if there is gravity. 
Only recently, Kim \& Lee~\cite{LK2022} acknowledged this crucial requirement for thermodynamics in general relativity. 
The authors insisted that the Tolman temperature gradient holds along directions perpendicular to both the particle trajectory and the heat flow:
\be{Tolman local}
\perp_a^c \mathcal{T}_c =0, \qquad
\perp_a^c \equiv \delta _a ^c + u_a u^c - \frac{q_a q^c}{q^2}.
\ee 
Note here that $\mathcal{T}_a = (\delta_a^c + u_a u^c) \mathcal{T}_c$ is a spatial vector normal to $u^a$, i.e., $\mathcal{T}_au^a =0$. 
The authors also reformulated the relativistic analogy of the Cattaneo equation to reflect the \textit{binormal equilibrium condition}~\eqref{Tolman local} by using the variational formulation of thermodynamics. 
The present article is based on the results rewritten for a steady heat flow state in Ref.~\cite{Kim:2023lta}.

\vspace{.2cm}
The theory of heat conduction consists of particle/entropy creation relations, two heat-flow equations, and the binormal equilibrium condition~\eqref{Tolman local}. The heat-flow equation comprises two differential equations: one is the relativistic analog of the Cattaneo equation, and the other originates from the $q^a$ part of the energy-momentum conservation equation.
In this work, we are interested in radial, steady heat flow.
Rather than describing all the details of the heat flow equations, we only present the steady-state equations developed in Ref.~\cite{Kim:2023lta}. In that work, the Landau-Lifshitz decomposition~\cite{LL,Tsumura:2012ss} with $v^a$ as a unit timelike vector was adopted so that the geometry is static to be consistent with the steady-state requirement.
The relation between the Landau-Lifschitz decomposition and the Eckart one is identified once we introduce a local Lorentz boost:
\be{u v}
v^a = \cosh \epsilon \, u^a + \sinh \epsilon \, \hat q^a, \qquad 
\hat j^a = \sinh \epsilon \, u^a + \cosh \epsilon \, \hat q^a,
\ee
where $\hat q^a \equiv q^a/q$, and $\epsilon$ is the boost parameter, respectively. 
$\hat j^a$ denotes the unit-spacelike vector along the heat flux normal to $v^a$ with $g_{ab} \hat j^a \hat j^b=1$.

\vspace{.3cm}
We are interested in a thermal system described by a generally static and spherically-symmetric geometry described by the metric,
\be{metric1}
ds^2 = -e^{2N} dt^2 + f^{-1} dr^2 + r^2 d\Omega^2_{(2)} ;
\qquad -g_{tt} = e^{2N} \equiv f(r) e^{-h(r)}, \qquad r_- \leq r \leq r_+,
\ee
where $d\Omega^2_{(2)}$ denotes the metric of a unit sphere.
Because we consider a steady state, all the metric functions are independent of time.
We assume that the thermal system is within a spherical shell from $r_- $ to $r_+$. 
In principle, engines (thermal baths) that transfer heat into matter or vice versa should be placed at both sides of the system.
In these coordinates,
$$
v^a \equiv (e^{-N}, 0, 0, 0), \qquad \hat j^a \equiv (0, \sqrt{f}, 0, 0).
$$
For an eternal black hole, the metric functions are $e^{2N} = f = 1-2M/r$ with $M$ being the Arnowitt-Deser-Misner mass of the black hole.

Now, let's briefly summarize the equations for steady radial heat flow, starting with the binormal components.
There are two binormal equations: the binormal equilibrium condition~\eqref{Tolman local} and the \textit{binormal part} of the relativistic analogy of the Cattaneo equation. 
Since heat flows along the radial direction, both equations exhibit simple angular independence:
\be{physicality}
\partial_k \Theta =0, \quad  \partial_k\mu =0, \qquad\mbox{where~} ~ k = 2,3.
\ee
Because $q$, $\Theta$, and $\chi$ are independent of angular coordinates, all physical quantities are also angularly independent. 
This result aligns with the spherical symmetry of the geometry.

\vspace{.1cm}
Four equations remain describing the behaviors along $u^a$ and $q^a$. 
In the context of a steady state of a thermal system, it has been argued that one of the four equations is redundant\footnote{See the last paragraph of Sec. 4 in Ref.~\cite{Kim:2023lta}.}.
Thus, we explicitly present three equations that describe the behaviors of the scalars $n$, $s$, and $q$. 
Two of these equations correspond to the particle conservation and the entropy creation equations:
\be{Gammas}
\Gamma_n = \nabla_a n^a =0, \qquad \Gamma_s = \nabla_a s^a = \frac{q^2}{\kappa \Theta^2},
\ee
respectively. 
Here, we choose the number creation rate to be zero. 
The first equation shows that
\be{id1}
J_\infty \equiv  -\sqrt{-g_{tt}} J(r) = - 4\pi r^2 \sqrt{-g_{tt}} n \sinh \epsilon  
\ee
is a position independent quantity, where $\hat j^a$ and $J(r)$ denote the unit vector along the radial direction in the metric~\eqref{metric1} and the total diffusion over a closed spherical surface, respectively.
The last equation comes from the $q^a$-parts of the relativistic analogy of the Cattaneo equation for the steady state with radial heat flow:
\be{r Tol2-2}
\left[ \log (e^N \Theta \cosh\epsilon)\right]' 
= -\frac{q}{\kappa \sqrt{f} \Theta\cosh\epsilon} 
	+ \frac{\vartheta}{\Theta} \tanh\epsilon 
	\left[ \log (\vartheta e^N \sinh\epsilon) \right]' .
\ee
Here the prime denotes the derivative with respect to $r$. 
This equation corresponds to Eq.~(56) in Ref.~\cite{Kim:2023lta}. 
In the case of a system evolving with time, an additional equation emerges from the heat part of the energy-momentum conservation relation.

We write the differential equations for steady states by using thermodynamic quantities explicitly in Sec.~\ref{sec:SHF}.
In Sec.~\ref{sec:bh}, we develop two approximations, which we use to analyze the thermodynamic system analytically. 
We then explicitly solve the steady state analytically and find exact analytic solution based on the mild-heat-flow approximation in Sec.~\ref{sec:IV}.
Then, we analyze the steady state equation in a less constrained approximation in Sec.~\ref{sec:V} and display solutions numerically and summarize the results in Sec.~\ref{sec:VI}.

\section{Steady heat flow in a spherically symmetric spacetime} \label{sec:SHF}

In this section, we analyze the equations for the steady thermal state in a spherically symmetric geometry undergoing radial heat flows.  
The formulation of the steady heat flow~\cite{Kim:2023lta} was done based on the Landau-Lifschitz decomposition, which corresponds to a kind of center of mass frame satisfying $- v^a \hat j^b T_{ab} =0$.
On the other hand, heat is defined in the Eckart decomposition in Eq.~\eqref{na}, based on the comoving observer with the matter.
The two decompositions are related by the local Lorentz boost~\eqref{u v}, where the boost parameter $\epsilon$ satisfies
\be{epsilon q}
\tanh 2\epsilon = \frac{q}{\varepsilon} ,
\qquad \varepsilon 
	= \frac12\left[ \rho +\Psi +\frac{\vartheta q}{\Theta}\right] ,
\ee
where $\rho+\Psi = n\chi + s \Theta= n\Theta(\mu+\sigma)$. 
Here $\mu \equiv \chi/\Theta$ and $\sigma \equiv s/n$ are the ratio of the chemical potential to temperature and the specific entropy, respectively.

\vspace{.1cm} 
The heat conduction equations~\eqref{Gammas},~\eqref{id1} and \eqref{r Tol2-2} are mixed together and form a coupled differential equation for $n$, $s$, and $q$.
Because we are interested in the behavior of the local temperature $\Theta$ rather than the entropy density $s$, we introduce a free energy 
\be{F}
\mathcal{F}(n, \Theta,\vartheta) \equiv \rho- \Theta s
\ee
and regard thermodynamic quantities as functions of $n$, $\Theta$, and $\vartheta$. 
Then, the first law of thermodynamics can be written as 
\bea
d \mathcal{F} &=& \chi dn  -s d\Theta + \varsigma d\vartheta 
= \left(\chi +\frac{q}{\Theta} \Big(\frac{\partial \vartheta}{\partial n} \Big)_{\Theta,q} \right) dn 
-\left( s -\frac{q}{\Theta}   
	\Big(\frac{\partial \vartheta}{\partial \Theta} \Big)_{n,q} 
	\right) d\Theta 
+ \frac{q}{\Theta} 
\Big(\frac{\partial \vartheta}{\partial q} \Big)_{n,\Theta} dq .
\label{dF}
\eea
With this form, the entropy $s$ is a function of $n$, $\Theta$, and $q$.
The specific entropy $\sigma = s/n$ can also be regarded as a function of them:
$$
d\sigma 
= \Big(\frac{\partial \sigma}{\partial n} \Big)_{\Theta,q}  dn 
+\Big(\frac{\partial \sigma}{\partial \Theta} \Big)_{n,q} d\Theta 
+\Big(\frac{\partial \sigma}{\partial q} \Big)_{n,\Theta}  dq .
$$

We now write the equations for a steady state one by one. 
\begin{enumerate}
\item
From Eq.~\eqref{id1}, we construct a function $\eta(r)$ of the metric functions: 
\be{eta r}
\eta(r) \equiv \frac{J_\infty}{4\pi r^2 \sqrt{-g_{tt}}} = \frac{J_\infty}{4\pi r^2 e^N}.
\ee

Because $\epsilon $ is a function of $q/\varepsilon$ as in Eq.~\eqref{epsilon q}, the particle number conservation equation~\eqref{id1} determines $q/\varepsilon$  as a function $\eta(r)/n$:
\be{id2}
n \sinh\epsilon =\eta(r).
\ee
This estabilishes the relationship:
$
q/\varepsilon = (2\eta/n) \sqrt{1+(\eta/n)^2}/(1+ 2(\eta/n)^2),
$
which expresses a connection between heat $q$ and total diffusion $-J_\infty$ at infinity.
Differentiating the equation~\eqref{id2} with respect to $r$, we get a differential relation,
\be{id3}
\epsilon' = \tanh\epsilon\left(\log \frac{\eta}{n}\right)'.
\ee
Note that the thermodynamic quantity directly related to $\epsilon$ is the number density $n$ only because $\eta$ is determined from the geometry~\eqref{eta r}.
Using the first equation in Eq.~\eqref{epsilon q} and interpreting $\varepsilon = \varepsilon(n,\Theta,q)$ as a function of $n$, $\Theta$, and $q$, we write Eq.~\eqref{id3} to the form:
\be{r:eq1}
A_1 \frac{n'}{n} + B_1 \frac{(e^N\Theta)'}{e^N\Theta} 
+C_1  \frac{q'}{q} = D_1,
\ee
where 
\bea
A_1 &\equiv& \frac{n}{\varepsilon} 
	\Big(\frac{\partial \varepsilon}{\partial n} \Big)_{\Theta,q}  
	- \sech^2\epsilon \sech 2\epsilon , \qquad
B_1 = \frac{\Theta}{\varepsilon}
	\Big(\frac{\partial \varepsilon}{\partial \Theta} \Big)_{n,q} ,
	\qquad
C_1 =	
\frac{q}{\varepsilon}  
\Big(\frac{\partial \varepsilon}{\partial q} \Big)_{n,\Theta} -1
, \nn \\ 
D_1 &\equiv &   -\sech^2 \epsilon \sech 2\epsilon \frac{\eta'}{\eta}
	 +\frac{N' \Theta} {\varepsilon}
	\Big(\frac{\partial \varepsilon}{\partial \Theta} \Big)_{n,q} .
\eea
Note that $A_1$, $B_1$, $C_1$, and $D_1$ contain terms without spatial derivatives of thermodynamic quantities.

\item
Next, we consider the second law of thermodynamics, the second equation in Eq.~\eqref{Gammas}. 
By using 
\be{sigma':1}
\nabla_a s^a = su^a \nabla_a \log \sigma +  \frac{1}{\sqrt{-g}} \partial_a \left( \sqrt{-g} \varsigma^a\right)  
= n \sinh\epsilon \sqrt{f}\sigma' +  \frac{q\sqrt{f} \cosh\epsilon}{\Theta} \left[ \log \frac{q \cosh\epsilon }{\eta \Theta}\right]',
\ee
the second law gives 
\be{sigma':31}
\left( \frac{q \cosh\epsilon }{\Theta} 
	-\eta \sigma\right)' 
=\frac{q^2}{\kappa\sqrt{f} \Theta^2}
	+\left( \frac{q\cosh\epsilon}{\Theta}
		- \eta \sigma \right)\frac{\eta'}{\eta}, 
\ee
where we use Eq.~\eqref{id3}. 
The equation can be written into the form:
\be{r:eq2}
A_2 \frac{n'}{n} + B_2 \frac{(e^N \Theta)'}{e^N \Theta} 
+ C_2 \frac{q'}{q} = D_2,
\ee
where 
\bea
A_2 &\equiv& - \frac{q}{\Theta} \sinh\epsilon \tanh\epsilon - n \eta  \Big(\frac{\partial \sigma}{\partial n} \Big)_{\Theta,q} , 
\qquad
B_2 \equiv - \frac{q\cosh\epsilon}{\Theta}
	- \eta \Theta 
	\Big(\frac{\partial \sigma}{\partial \Theta} \Big)_{n,q} 
,
\nn \\
C_2 &\equiv &  \frac{q\cosh\epsilon}{\Theta}
	- \eta q \Big(\frac{\partial \sigma}{\partial q} \Big)_{n,\Theta} ,
\qquad
D_2 =  \frac{q^2}{\kappa\sqrt{f} \Theta^2}
	+ \frac{q}{\Theta \cosh\epsilon} \frac{\eta'}{\eta}
	+B_2 N' .
\eea

\item  We next consider the $q^a$-part of the relativistic analogy of the Cattaneo equation~\eqref{r Tol2-2}. 
Using Eq.~\eqref{id3}, and interpreting $\vartheta$ as a function of $n$, $\Theta$, and $q$, we rewrite Eq.~\eqref{r Tol2-2} into the form:
\be{r:eq3}
A_3 \frac{n'}{n} + B_3 \frac{(e^N \Theta)'}{e^N \Theta} 
+ C_3 \frac{q'}{q} = D_3,
\ee
where 
\bea
A_3 &\equiv&\frac{\vartheta}{\Theta} \tanh\epsilon 
	- \tanh^2\epsilon 
	-\frac{\vartheta \tanh\epsilon}{\Theta } 
		\frac{n}{\vartheta} \Big(\frac{\partial \vartheta }{\partial n} \Big)_{\Theta,q} , 
\nn \\
B_3 &\equiv& 1 -\frac{\vartheta \tanh\epsilon}{\Theta}
	\frac{\Theta}{\vartheta} 
	\Big(\frac{\partial \vartheta}{\partial \Theta} \Big)_{n,q} 
, \qquad
C_3 \equiv -\frac{ \vartheta  \tanh\epsilon }{\Theta}
	\frac{q}{\vartheta} \Big(\frac{\partial \vartheta}{\partial q} \Big)_{n,\Theta} ,
\nn \\
D_3 &=& -\frac{q}{\kappa \Theta\cosh\epsilon \sqrt{f}} 
+ \left(1 -\frac{\Theta}{\vartheta} 
	\Big(\frac{\partial \vartheta}{\partial \Theta} \Big)_{n,q} 
	 \right)\frac{\vartheta \tanh\epsilon }{\Theta} N'  
-\tanh\epsilon \left[\tanh\epsilon
-\frac{\vartheta}{\Theta} \right] \frac{\eta'}{\eta} 
  .
\eea

\end{enumerate}

Combining Eqs.~\eqref{r:eq1}, \eqref{r:eq2}, and \eqref{r:eq3}, we get how $n$ and $e^N\Theta$ varies spatially:
\bea \label{sol}  
\frac{ n'}{n} &=&  -\frac{1}{D} 
		\left[D_1 (B_2 C_3- B_3 C_2)
		+D_2(B_3 C_1- B_1 C_3) +D_3(B_1C_2-B_2C_1) \right],  \nn \\
\frac{(e^N \Theta)'}{e^N \Theta} & =&  -\frac{1}{D} 
		\left[D_1 (A_3 C_2- A_2 C_3)
		+D_2(A_1 C_3- A_ 3C_1) 
			+D_3(A_2C_1-A_1C_2) \right], 
\eea
where 
\be{D}
D \equiv 	A_3 B_2 C_1-A_2 B_3
   C_1-A_3 B_1 C_2+A_1 B_3 C_2+A_2
   B_1 C_3-A_1 B_2 C_3.
\ee
For the case of the differential of heat, $q'$, Eq.~\eqref{id2} plays its integrated form. 
Therefore, we do not need to add another differential equation.
Later in this work, we concentrate on solving the differential equations and understanding the implication of $\vartheta$ on $q$ on the heat flow.

\section{Two approximations for thermal systems in steady heat flow  } \label{sec:bh}

In this section, we introduce two approximations which help us analyze and solve the steady heat flow analytically. 

\subsection{Low-boost approximation}
We begin by considering Eq.~\eqref{epsilon q} utlizing the Schwarz inequality:
\be{approx1}
\tanh 2|\epsilon| \equiv \frac{2|q|}{n\Theta(\mu+\sigma) + \vartheta q/\Theta} 
= \frac{2}{ \frac{n \Theta (\mu+\sigma)}{|q|}
	+ n \beta (\mu+\sigma) \frac{|q|}{n\Theta (\mu+\sigma)}}  \leq \frac1{\sqrt{n \beta (\mu+\sigma)}}.
\ee 
This outcome suggests that the boost parameter $|\epsilon|$ has an upper bound determined by $n\beta(\mu+\sigma)$ where $\beta =\vartheta/q$. 
When $n\beta(\mu+\sigma)$ is large enough, the boost parameter can be small enough irrespective of heat. 
Therefore, we first consider the \textit{low-boost approximation},
\be{approx1}
|\epsilon|  \approx \frac{|q|}{n\Theta(\mu+\sigma) + \vartheta q/\Theta} \ll 1.
\ee 
Here, the symbol $\approx$ denotes that the equality holds under the low-boost approximation. 
For ordinary matter, satisfying $|q| \leq n \sigma \Theta$ due to the time-likeness of the vector $s^a$, the value of $|q|$ may not exceed the first term in the denominator. 
On the other hand, for certain cases, such as when $\mu+\sigma=0$ (dark energy) or $\Theta \sim 0$, the second term dominates the denominator. 
Even in such cases, the low boost approximation could be valid. 
In this work, we concentrate on ordinary matters that satisfies $\vartheta q/\Theta \ll n \Theta(\mu+\sigma)$. 
This condition allows the low-boost approximation to take the form:
\be{approx2}
\epsilon  \approx \frac{q}{n\Theta(\mu+\sigma)} +O(\epsilon^2)  .
\ee
Here we assume $\vartheta \sim O(1)$ rather than $\sim O(\epsilon)$.
Consequently, we treat $\vartheta q/\Theta$ as a first-order term in the approximation.

The equation~\eqref{approx1} constrains the number density, in combination with the number conservation equation~\eqref{id1} and \eqref{eta r},
\be{n mild}
|\epsilon| \approx \frac{|\eta(r)|}{n} = \frac{|J_\infty|}{4\pi r^2 e^N n}\ll 1 
\quad \rightarrow \quad n \gg \frac{|J_\infty|}{4\pi r^2 e^N}.
\ee
Therefore, there should exist large enough number of particles to support the diffusion.
For typical values of $n$, this condition implies $|J_\infty| \ll 1$. 
If this requirement is to be met near the event horizon, it demands $n \to \infty$ given that $-g_{tt} = e^{2N} \to 0$ in that vicinity. Therefore, we should be cautious when applying the low boost approximation in the vicinity of an event horizon.

\vspace{.2cm}
In this section, we consider a general thermal system formally without introducing an explicit form of the master function $\Lambda$.
To solve the evolution equation~\eqref{sol}, it is necessary to express explicitly the determinant $D$ and the other coefficients $A_k$, $B_k$, $C_k$ and $D_k$ in terms of thermodynamic quantities.
With the low-boost approximation, $A_k$, $B_k$, $C_k$ and $D_k$ become:
\be{ABCD}
\left(
\begin{tabular}{ccc}
$A_1$ & $A_2$ & $A_3$    \\
$B_1$ & $B_2$ & $B_3$    \\
$C_1$ & $C_2$ &  $C_3$    \\
$D_1$ & $D_2$ & $D_3$   
\end{tabular}
\right)  \approx
\left(
\begin{tabular}{ccc}
$\displaystyle \frac{n}{\varepsilon} 
	\Big(\frac{\partial \varepsilon}{\partial n} \Big)_{\Theta,q}  
	-1$ & 
 $\displaystyle  - n \eta  \Big(\frac{\partial \sigma}{\partial n} \Big)_{\Theta,q} $ & 
 $\displaystyle  \frac{\vartheta \epsilon}{\Theta}
	 \left[1- \frac{n}{\vartheta} \Big(\frac{\partial \vartheta}{\partial n}\Big)_{\Theta,q}\right]$    \\
$\displaystyle \frac{\Theta}{\varepsilon}
	\Big(\frac{\partial \varepsilon}{\partial \Theta} \Big)_{n,q} $ & $\displaystyle - \frac{q}{\Theta}- \eta \Theta 
	\Big(\frac{\partial \sigma}{\partial \Theta} \Big)_{n,q} $ & $\displaystyle  1 -\left(\frac{\vartheta \epsilon}{\Theta} \right) 
	\frac{\Theta}{\vartheta} \Big(\frac{\partial \vartheta}{\partial \Theta}\Big)_{n,q} $    \\
$\displaystyle \frac{q}{\varepsilon}  
\Big(\frac{\partial \varepsilon}{\partial q} \Big)_{n,\Theta} -1$ & $\displaystyle \frac{q}{\Theta} -\eta q \Big(\frac{\partial \sigma}{\partial q} \Big)_{n,\Theta}$ &  $\displaystyle  -\left(\frac{\vartheta \epsilon}{\Theta} \right) 
	\frac{q}{\vartheta} \Big(\frac{\partial \vartheta}{\partial q}\Big)_{n,\Theta} $    \\
$\displaystyle   - \frac{\eta'}{\eta}
	 +\frac{N' \Theta} {\varepsilon}
	\Big(\frac{\partial \varepsilon}{\partial \Theta} \Big)_{n,q} $ & $\displaystyle  \frac{q}{\Theta} \frac{\eta'}{\eta}
	+B_2 N' $ & $\displaystyle
	\frac{\vartheta \epsilon}{\Theta} \left[\left(1- \frac{\Theta}{\vartheta} \Big(\frac{\partial \vartheta}{\partial \Theta}\Big)_{n,q}\right)N'+ \frac{\eta'}{\eta} \right] -\frac{q}{\kappa \Theta \sqrt{f}} $   
\end{tabular}
\right)  .
\ee
where we can use $N' + \eta'/\eta = - 2/r$ to replace $\eta'/\eta$ with $-2/r -N'$  from Eq.~\eqref{eta r}. 
Even though the equation of motion~\eqref{sol} does not allow an exact analytic solution,
as we will show in Sec.~\ref{sec:V}, this approximation enable us to analyze the behavior of the solution analytically, even for non-perturbative values of $\vartheta$. 

\subsection{Mild-heat flow approximation}
To find analytic solutions, we further assume that the heat flows mildly enough to ignore terms of $O(q^2)$.
In this case, the boost parameter $\epsilon$ automatically becomes small enough for ordinary matter satisfying $\mu+\sigma \neq 0$.
Note also that $\rho$ is an even function\footnote{From the comment just before Eq.~\eqref{1st law}, the $q$ dependence can be deduced from the term $d\rho \sim \cdots+\varsigma d\vartheta$, where $\varsigma = q/\Theta$ and $\vartheta=\beta q$. In addition, if the energy density has a linear term in $q$, the thermal equilibrium state cannot be stable under the perturbation of heat.} of $q$.
Then, we expect that $\frac{q}{\varepsilon} \left(\frac{\partial \varepsilon}{\partial q} \right)_{n,\Theta}$, $\frac{q}{\vartheta} \Big(\frac{\partial \vartheta}{\partial q}\Big)_{n,\Theta}$, and $q \Big(\frac{\partial \sigma}{\partial q} \Big)_{n,\Theta}$ must be even functions of $q$ being $O(q^2)$ at least.
Therefore, without loss of generality, we may set $C_1 \simeq -1$, $C_2 \simeq q/\Theta$, and $C_3 \simeq 0$, where the symbol `$\simeq$' denotes that we use the mild-heat flow approximation.
This condition also implies that 
\be{approx}
\vartheta \simeq O(q)
\ee
and we ignore terms containing $\vartheta \epsilon$ because they are $O(q^2)$.
Based on this simplification, we find an exact steady heat-flowing solution for a system consisting of an ideal gas in the next section.

This mild heat-flow approximation simplifies one of the equation of motions~\eqref{sol} enough to analyze the results without explicit form of $\Lambda$.
The determinant $D $ in Eq.~\eqref{D} becomes:	
\be{D:mild}
D \simeq -A_2C_1 +A_1 C_2 =  
\frac{q}{\Theta}\left[ \frac{n}{\varepsilon} 
	\Big(\frac{\partial \varepsilon}{\partial n} \Big)_{\Theta,q}  
	-1 
	-\frac{n\Theta\eta}{q} 
	    \left(\frac{\partial \sigma}{\partial n}\right)_{\Theta,q}
	\right] .
\ee
The first equation of Eq.~\eqref{sol} gives  
\bea
\frac{ n'}{n} & \simeq &  -\frac{1}{D} 
		\left[C_1D_2- C_2D_1 +D_3(B_1C_2-B_2C_1) \right]
\approx  
\frac{ 
	1 -\frac{\Theta}{\varepsilon} \left(\frac{\partial \varepsilon}{\partial \Theta} \right)_{n,q}
	+\frac{\eta \Theta^2}{q}  
		\left(\frac{\partial \sigma}{\partial \Theta} \right)_{n,q} 
	}{
	1- \frac{n}{\varepsilon} 
	\left(\frac{\partial \varepsilon}{\partial n} \right)_{\Theta,q}  
	 +
	\frac{\eta\Theta n}{q} \left(\frac{\partial \sigma}{\partial n}\right)_{\Theta,q}
	} N' .
	\label{n':mh}
\eea 
Note that $n'/n \propto N'=(\log \sqrt{-g_{tt}})'$.
This result clearly signifies that the density variation has a geometric origin even though the details are affected by the thermodynamic properties.
The second equation presents the relation satisfied by the local temperature,
\be{Theta'}
\frac{(e^N\Theta) '}{e^N \Theta} \simeq  D_3 = -\frac{q}{\kappa \Theta \sqrt{f} } 
 = - \frac{J_\infty (\mu+\sigma)}{4\pi \kappa r^2 \sqrt{f}e^N} ,
\ee
where we have used  Eq.~\eqref{approx2} in the last equality.
This temperature equation was already noticed in Eq.~(63) in Ref.~\cite{Kim:2023lta}.
The local temperature behaves as
\be{T mild}
\frac{\Theta'}{\Theta} =-N'  -
	\frac{(\mu+\sigma) J_\infty}{4\pi \kappa f \sqrt{-g} } ,
\ee
where the first/second term in the right-hand side has a geometric/thermodynamic origin, respectively. 
When $q \to 0$ ($J_\infty \to 0$), this formula reproduces the Tolman's temperature relation~\eqref{Tolman1}.  
An interesting observation here is that both $N'$ and $1/f$ diverges as $r \to 2M$ with the same way.
When $q=0$, the right-hand side vanishes and the local temperature becomes divergent evidently because of the $N'$ term.
On the other hand, in the presence of a heat, the other possibility happens.
When $J_\infty$ satisfies
$$
J_\infty = -\frac{4\pi \kappa M}{(\mu + \sigma)_{r=2M}},  
$$
the spatial derivative of the local temperatur $\Theta'$ goes to zero at the horizon making the local temperature finite because 
$$
 \lim_{r\to 2M}\left[ -4\pi \kappa \frac{N'f \sqrt{-g}}{ \mu + \sigma} \right] = -\frac{4\pi \kappa M}{(\mu + \sigma)_{r=2M}}  = J_\infty.
$$
Here we use the near horizon limit, $\lim_{r\to 2M} N' f = 1/4M$. 
If this possibility is right, the heat behaves as, from Eqs.~\eqref{approx2} and \eqref{n mild},
\be{q:ig}
q \approx \frac{J_\infty \Theta (\mu+\sigma)}{4\pi r^2 e^N}
 = -\frac{\kappa M}{4\pi r^2 e^N} \frac{\Theta (\mu+\sigma)}{(\mu + \sigma)_{r=2M}}, 
\ee
At this point, we need to check the low-boost approximation~\eqref{approx2}.
By using Eqs.~\eqref{n':mh} and \eqref{q:ig}, the approximation gives the inequality
$$
|\epsilon| \simeq \frac{|q|}{n(\mu+\sigma)\Theta}
 \approx  \frac{\kappa M}{4\pi r^2 e^N} \frac{1}{n(\mu + \sigma)_{r=2M}} \ll 1.
$$
This inequality will hold if $n e^N \gg 1$ as $r \to 2M$. 
From Eq.~\eqref{n':mh}, we will get $n \propto e^{\alpha N}$ where $\alpha$ will be determined from the near-horizon behavior of the thermodynamic quantities in the coefficient of $N'$ in the equation~\eqref{n':mh}.
Therefore, the validity of the approximation is determined by the model dependent value $\alpha$ relative to one.
Therefore, we need to examine the situation in detail by using an explicit example.

\section{Steady heat flow of an ideal gas around a black hole}
\label{sec:IV}
In this section, we introduce an ideal gas with heat flow and investigate systems with steady heat flow in a spherically symmetric geometry described by the metric~\eqref{metric1}. 
Initially, we consider a general, spherically symmetric geometry, allowing us to take into account the back-reaction of matter on the geometry through the Einstein equation formally. 
Later, we solve the equations for a steady thermal system in a background Schwarzschild black hole.

\subsection{Equation of motions for an ideal gas with heat flow under low-boost approximation} \label{sec:ig}
As an explicit model of matter consisting a thermodynamic system, we consider the ideal gas system developed in Ref.~\cite{Kim:2023wel} with its energy density having dependence on heat. 
The energy density of the ideal gas takes the form,
\be{rho:ig}
\rho(n, s, \vartheta) = n c_v \Theta + m n.
\ee
where the heat dependence of the energy density comes from the temperature indirectly,
\be{Theta}
\Theta(n, \sigma, \vartheta) \equiv \Theta_0 \left(\frac{n}{n_0}\right)^{1/c_v}  \exp \left[\Phi(\vartheta)+ \frac{\sigma}{c_v}\right].
\ee
The function $\Phi(\vartheta)$ is known to have a minimum value zero at $\vartheta =0$, expanding around the minimum value,   
\be{q:ig1}
\Phi = \gamma \frac{\vartheta^2}{\vartheta_0^2} +\cdots ,\qquad 
\Phi'(\vartheta) =\frac{q}{ c_v n \Theta^2}  .
\ee
where the second equation comes from the first law~\eqref{1st law}.
In general, $\Phi'' \geq 0$.
From the last equation, $\vartheta(Q)$ was shown to be a function of $Q$ only with $\vartheta(0)=0$ where
 \be{Q:def}
Q \equiv \frac{\vartheta_0 q}{\gamma c_v n \Theta^2}.
\ee

 We interpret the specific entropy $\sigma$ as a function of $n$, $\Theta$, and $\vartheta$ from Eq.~\eqref{Theta} with the form:
\be{sigma:ig}
\sigma(n, \Theta, q) = c_v\log \frac{\Theta}{\Theta_0} - \log \frac{n}{n_0} - c_v \Phi(\vartheta).
\ee
For later convenience, we write the partial derivatives of $\vartheta$ and $\sigma$:
\bea \label{dvT:dn}
&&n\left(\frac{\partial \log \vartheta}{\partial n}\right)_{\Theta,q} =- \frac{\Phi'}{\vartheta \Phi''}, 
\qquad
\Theta\left(\frac{\partial \log \vartheta}{\partial \Theta}\right)_{n,q} =- \frac{2\Phi'}{\vartheta \Phi''},
\qquad
q\left(\frac{\partial \log \vartheta}{\partial q}\right)_{n,\Theta} =\frac{\Phi'}{\vartheta \Phi''}, \nn \\
&& n\left(\frac{\partial \sigma}{\partial n}\right)_{\Theta,q}
	=-1 +c_v \frac{(\Phi')^2}{\Phi''}, 
\qquad 
\Theta \left(\frac{\partial \sigma}{\partial \Theta}\right)_{n,q} 
	= c_v\left(1  + 2\frac{(\Phi')^2}{\Phi''} \right), 
\qquad
q\left(\frac{\partial \sigma}{\partial q}\right)_{n,\Theta}
	=-c_v \frac{(\Phi')^2}{\Phi''} .
\eea
Note that all of these derivatives are just a function of $Q$ only.
Note also that the specific heat for constant volume and heat,  $c_{v,q} \equiv \Theta \left(\frac{\partial \sigma}{\partial \Theta}\right)_{n,q}$,  gains correction term proportional to $(\Phi')^2/\Phi''$ from the heat $q$.
We further calculate the partial derivatives of $\varepsilon$ with respect to $n$, $\Theta$, and $q$, 
\bea
\frac{n}{\varepsilon}\left(\frac{\partial \varepsilon}{\partial n}\right)_{\Theta,q} &=& 
1- \frac{\vartheta q}{2\varepsilon \Theta} \left(1+\frac{\Phi'}{\vartheta \Phi''}\right), 
\nn \\
\frac{\Theta}{\varepsilon}\left(\frac{\partial \varepsilon}{\partial \Theta}\right)_{n,q} 
&=&  \frac{n\Theta(1+c_v)}{2\varepsilon}
	 - \frac{\vartheta q}{2\varepsilon\Theta} \left(1+ 2\frac{\Phi'}{\vartheta \Phi''} \right),
\nn \\
\frac{q}{\varepsilon}
	\left(\frac{\partial \varepsilon}{\partial q}\right)_{n,\Theta} 
	&=& 
	\frac{\vartheta q}{2\varepsilon \Theta} \left(1+\frac{\Phi'}{\vartheta \Phi''}\right) .  \label{dlog ve}
\eea
Here, we use $
\varepsilon  =\frac{mn}{2}+ \frac{c_v+1}2 n\Theta  + \frac{\vartheta q}{2\Theta}
$ from Eqs.~\eqref{rho:ig} and Eq.~\eqref{epsilon q}.

\vspace{.2cm}
Now we write the heat-flow equation of motions~\eqref{sol} for the  ideal gas in a background Schwarzschild geometry explicitly based on the low-boost approximation~\eqref{approx1}.
Here, we use Eqs.~\eqref{dvT:dn} and \eqref{dlog ve}.
From Eqs.~\eqref{D} and \eqref{n mild}, we have 
$
D \approx 
\left[1- c_vF\right] \eta   ,
$ where $F \equiv \Phi'^2/\Phi''$.
The equations in Eq.~\eqref{sol} become
\bea
\frac{n'}{n} & \approx & 
	\frac{c_v F}{ (c_v
   F-1)} \left(N'+\frac{\eta'}{\eta}\right)
   +\frac{\frac{m  c_v F}{(1+c_v)\Theta +m}+c_v +\frac{m}{\Theta} }{c_v F-1} N',  \label{eom1} \\
\frac{(e^N\Theta)'}{e^N\Theta} & \approx & 
-\frac{ \eta (1+c_v+m/\Theta)}{\sqrt{f}   \kappa }
-\frac{\gamma c_v}{1+ c_v + \frac{m}{\Theta}}\frac{Q\vartheta}{\vartheta_0}
 \left\{
	\frac{B+c_v F}{c_v F-1} \frac{m N'}{(1+c_v)\Theta +m} 
	+\frac{B+1}{c_v F-1} 
		\left(
			\frac{ q}{\eta  \Theta}N'
		+\frac{\eta'}{\eta }\right) \right\} , 
\nn
\eea	
where $B \equiv \frac{\Phi'}{\vartheta \Phi''}$. 
Here we use $q/\eta \Theta \approx q/\epsilon n \Theta =\mu+\sigma=1+ c_v + m/\Theta$ from Eqs.~\eqref{eta r}, \eqref{id2}, \eqref{approx} and $\rho+\Psi= n\Theta(\mu+\sigma)$.
Until now, we use the low-boost approximation~\eqref{approx1} only and did not adopt the mild-heat flow approximation~\eqref{approx}.
In the calculation of the second line, we also use
$$
\frac{\eta  \vartheta}{n\Theta } = \frac{\eta \Theta}{q} 
	 \frac{\vartheta_0 q}{\gamma c_v n\Theta^2} \frac{\vartheta \gamma c_v}{\vartheta_0 } 
\approx  \frac{\gamma c_v}{1+ c_v + \frac{m}{\Theta}}\frac{Q\vartheta}{\vartheta_0}
$$
Note that $(Q \vartheta/\vartheta_0)$, $B$, and $F$ are functions of $Q$ only.

\subsection{Exact steady state solution of an Ideal gas with mild-heat flow approximation}\label{sec:4B}
In this subsection, we employ the mild-heat-flow approximation to find heat-flowing solutions in a background Schwarzschild geometry.
Therefore, we ignore the back-reaction of the thermodynamic matter to geometry.
We examine the mild-heat flow approximation, which requires the value of $Q$ to satisfy
\be{Q << 1}
|Q|= \left( \frac{m+\Theta(c_v+1)}{ n \Theta^2}\right) \left(\frac{ |\vartheta_0 J_\infty|}{4\pi \gamma c_v r^2 e^N} \right) 
= \frac{\vartheta_0}{\gamma c_v\Theta} \left(\frac{m}{\Theta}+c_v+1\right) |\epsilon|  \ll 1. 
\ee
Here, $\gamma$ and $\vartheta_0$ are theory-dependent constants in Eq.~\eqref{q:ig1}, and $J_\infty$ is a position-independent quantity that determines the strength of heat flux.
Then, the quantity $\vartheta$ in Eq.~\eqref{q:ig1} becomes linear in $Q$:
$$
\frac{\vartheta}{\vartheta_0} = Q + O(Q^2) ,
$$
making $\vartheta $ to be $O(q)$.
Eventually, this result makes the steady state to be independent of the second order terms having $\gamma$.
Based on the approximation, we simplify Eq.~\eqref{eom1} by using this result and $B-1 \sim O(Q^2)$ to get
\bea
\frac{n'}{n}&\approx & 
 -  N' \left( \frac {m}{\Theta} + c_v  \right) , \nn \\
\frac{\Theta'}{\Theta} 
&=& - N' - \frac{J_\infty}{4\pi \kappa} \frac{  \frac{m}{\Theta} + c_v +1}{f \sqrt{-g} } 
= - \left(N' + \frac{J_\infty (1+c_v)}{4\pi \kappa f\sqrt{-g}} \right) 
	-\frac{J_\infty }{4\pi \kappa f\sqrt{-g}}  \frac{m}{\Theta} .
\label{T mild2}
\eea
We have also removed $q$ by using Eqs.~\eqref{approx2} and \eqref{n mild} to set 
\be{q:1}
q = \frac{J_\infty[ (1+c_v)\Theta + m]}{4\pi r^2e^N}.
\ee
Note that this differential equations~\eqref{T mild2} are determined from the thermal equilibrium configuration only and are independent of the higher order dependence in $q$.

\vspace{.3cm}
Until now, we did not fix the geometry but have used the general form in Eq.~\eqref{metric1} so that the matter can determine geometry through Einstein equation.
Now, we require the geometry to be given by an eternal Schwarzschild black hole, 
\be{metric:Nf}
e^{2N} = f = 1- \frac{2M}{r} ,
\ee
where $M$ is the Arnowitt-Deser-Misner mass of the black hole.
Here, we consider a system located outside the black hole, and heat flows out of or into the black hole depending on the signature of the heat.
We assume that the reaction of the matter to the geometry is negligible and treat the geometry as a background.
Then, $N' = M/(r(r-2M))$ and $\sqrt{-g} = e^{N} f^{-1/2} r^2 = r^2$.
Here, we assume that the system is located at the equatorial plane without loss of generality because of spherical symmetry.
 Then, the two equations in Eq.~\eqref{T mild2} become
\be{two eq}
\frac{d \log n}{dy} 
	= -  \frac{\frac {m}{\Theta} + c_v}{2y(y-1) }  , \qquad
\frac{d\log(\Theta-\Theta_{H} )}{dy} =  -\frac{1+j}{2y(y-1)}, 
\ee
where $y = r/2M$ and 
\be{aT}
j \equiv  \frac{(1+c_v)J_\infty}{4\pi \kappa M}
 \qquad
\Theta_H \equiv -\frac{m j}{(1+c_v) (1+j)} .
\ee
The signature of $\Theta_H$ is negative when $j > 0$ or $j < -1$.
It is positive when $-1 < j < 0$.
Note also that $\Theta_H =0$ when $m=0$ or $j =0$.

\subsubsection{Exact solution for thermal system with massless particles} \label{sec:4B1}

When the mass of the particle goes to zero, $m=0$, the solution to Eq.~\eqref{two eq} takes a simple form. 
Therefore, we first consider the massless case, which gives $\Theta_H =0$. 
The two differential equations in Eq.~\eqref{two eq} present an exact solution,
\be{massless}
\Theta = T_\infty \left(1-\frac{2M}{r} \right)^{-\frac{1+j}2}, \qquad n =n_0 \left(1-\frac{2M}{r} \right)^{-\frac{c_v}2},
\ee
where $n_0$ and $T_\infty$ are integration constants denoting the asymptotic value of $n$ and the local temperature, respectively.
From Eq.~\eqref{q:1}, the heat behaves as
\be{q:m=0}
q =  \frac{j\kappa M T_\infty}{r^2}\left(1- \frac{2M}r\right)^{-\frac{j+2}2} . 
\ee
Let us check the low-boost approximation by examine the value
$$
|\epsilon| \approx \frac{|q|}{n\Theta (\mu+\sigma)} 
= \frac{|J_\infty|}{ 4\pi n_0r^2} \left(1-\frac{2M}{r}\right)^{\frac{c_v-1}2} \ll 1.
$$
This inequality holds for all regions outside the event horizon when $c_v \geq 1$ and $|J_\infty|$ is small enough. When the low-boost approximation is valid, the mild-heat flow approximation also holds when 
\be{Q:ig}
|Q|  = \left( \frac{\kappa M}{ n \Theta}\right) \left(\frac{ |\vartheta_0 j|}{\gamma c_v r^2 e^N} \right) 
=\left(\frac{ \kappa M|\vartheta_0 j|}{\gamma c_v n_0 T_\infty } \right) 
	\frac1{r^2}\left(1-\frac{2M}r\right)^{\frac{c_v+j}2} \ll 1.
\ee
This result clearly shows that the approximation fails to hold near the horizon when $j< - c_v $. In this case, the approximation will be valid only for the region $r> r_c~ ( > 2M)$, where $r_c$ is the radius satisfying $|Q(r_c)| \sim 1$. On the other hand, the approximation holds for all the regions outside the horizon when 
\be{a mild}
j \geq -c_v .
\ee
Let us examine the results one by one.
\begin{itemize}
\item The asymptotic temperature is $T_\infty$, which must be non-negative. The local temperature will monotonically increase/decrease from zero/infinity to the asymptotic temperature when $j \lessgtr -1$, respectively.

\item The number density of the particle monotonically decreases from infinity at the horizon to its asymptotic value irrespective of the heat flow once $c_v$ is positive. 

\item When $j=0$ (thermal equilibrium), the heat vanishes. When $j< -2$ (heat flows in fast), the heat goes to zero at the horizon. This is an interesting possibility; however, we should be cautious in applying this result because the mild-heat flow approximation is invalid at the horizon when $j< -c_v$~\eqref{a mild}. Therefore, only when $c_v> 2$, this possibility could be achieved. When $j> -2$, the value of the heat diverges at the horizon.

\item Specifically when $j=-1$ (heat flow in.), the local temperature takes a position-independent, finite value for the massless particles.
\end{itemize}

\subsubsection{Exact solution for thermal system with massive particles} \label{sec:4B2}

Even when the mass of particles does not vanish, $m\neq 0$, we can solve the heat-flow equations~\eqref{two eq} exactly.  

In this subsection, we consider the solution to Eq.~\eqref{two eq} when $m\neq 0$.
Integrating the second equation in Eq.~\eqref{two eq}, we find  the local temperature
\be{Tr}
\Theta = 
\left\{
\begin{array}{cc}
\displaystyle \Theta_H+ (T_\infty-\Theta_H)\left(1- \frac{2M}r\right)^{-\frac{1+j}2 },   &   
j \neq -1, \vspace{.1cm}  \\
\displaystyle T_\infty+ \frac{m}{2(1+c_v)} \log \left(1- \frac{2M}{r}\right) ,  & j=-1 ,   \\   
\end{array}
\right. 
\ee
where $T_\infty $ is an integration constant denoting the asymptotic temperature.
Therefore, $T_\infty \geq 0$ should be satisfied for the asymptotic temperature be non-negative.
Let us analyze the results case by case:
\begin{itemize}
\item 
When $T_\infty = \Theta_H$, the local temperature is homogeneous irrespective of $j$. 

\item
$j< -1$  ($\Theta_H< 0$) case: \\
The temperature increases monotonically from $\Theta_H$ at the horizon to $T_\infty$ as $r \to \infty$.
In this case, there exists a radius $r_0$ outside the horizon where the temperature goes to zero:
\be{r0:ig}
r_0 = \frac{2M}{1- (1-T_\infty/\Theta_H)^{\frac{2}{1+j}}}  \geq 2M .
\ee
Inside this surface, the local temperature appears to have unphysical negative value. 
This is because we have used the mild-heat flow approximation, which may not be valid for $r \lesssim r_0$.

\item $j=-1$ case: \\
The local temperature monotonically increases from negative infinity at the horizon to $T_\infty$ asymptotically.
It vanishes at a surface $r=r_0$ outside the horizon,
$$
r_0 = \frac{2M}{1- e^{-2(1+c_v)T_\infty/m}} > 2M .
$$
The applicability of the mild-heat flow approximation is questionable in the region $r \lesssim r_0$.

\item $-1 < j < 0$ ( $\Theta_H>0$) case: \\
When $T_\infty \gtrless \Theta_H$, the local temperature goes to positive/negative infinity as $r\to 2M$.
Depending on the signature, the local temperature gradually decreases/increases and approaches the asymptotic value with $r$.
Therefore, when $T_\infty<\Theta_H$, there exists a surface of vanishing local temperature at the outside of the horizon.
The radius of the surface is formally given by the same form as the formula~\eqref{r0:ig}.
When $T_\infty> \Theta_H$, the local temperature monotonically decreases from infinity to the asymptotic value. 
Therefore, there does not exist the surface of vanishing temperature. 

\item $j=0$ ($\Theta_H =0$) case:\\
The asymptotic temperature is nothing but $T_\infty$ itself, which should be non-negative.
Because $J_\infty =0$, heat does not flow.
Therefore, this state corresponds to the thermal equilibrium.

\item $j>0$  ($\Theta_H < 0$) case:\\
The local temperature diverges at the horizon and monotonically decreases to the asymptotic temperature with $r$. 
Heat flows out.
\end{itemize}

We integrate the first equation in Eq.~\eqref{two eq} by using the solution~\eqref{Tr}.
After the change of variable $1-y^{-1} = e^z$, we get
$$
\log n 
	= - \frac{c_v}{2}\int dz \left[1 + \frac{m / (c_v(T_\infty-\Theta_H) )}
		{\Theta_H /(T_\infty-\Theta_H) + e^{-(1+j)z/2} } \right] .
$$
The integration in the right-hand side presents distinguished forms of solution depending on $j= -1, 0 $ or else.
Explicitly we get
\be{nr}
n =n_0 \left(1- \frac{2M}{r} \right)^{-\frac{c_v}2 }  \times
\left\{
\begin{array}{cc}
\displaystyle  
	\left[1+ \frac{\Theta_H}{T_\infty} \left(1-\frac{2M}{r}\right)^{\frac{1+j}2}
		-\frac{\Theta_H}{T_\infty} \right]^{\frac{1+c_v}{j}} ,   &   
j\neq -1, 0,  \vspace{.1cm}  \\
\displaystyle 
	\left[ 1+\frac{m}{2 T_\infty (1+c_v)} 
	\log \left(1- \frac{2M}{r} \right)\right]^{-(1+c_v)} ,  & j=-1  , 
	\vspace{.1cm}   \\ 
\displaystyle 
\exp \left[ -\frac{m}{ T_\infty} \left(1-\frac{2M}{r}\right)^{1/2} + \frac{m}{T_\infty} \right] ,  & \displaystyle 
j=0 ,    \\   
\end{array}
\right.
\ee
where $n_0$ is an integration constant denoting the asymptotic value\footnote{The asymptotic value may hold when the system is stable from the horizon to the asymptotic region. In a relativistic thermodynamic system, there could exist other instabilities such as the Jean's instability~\cite{Jeans} for such a large size system. Therefore, we simply regard the value as one of the constants representing the thermodynamic system. }.
Let us observe the behaviors of the number density for each case.
\begin{itemize}

\item 
$j \leq -1$ ($\Theta_H < 0$) case: \\
The number density becomes divergent at $r_0 (> 2M)$. 
The mild-heat-flow approximation holds only for the region outside the surface, $r> r_0$.

\item 
$-1 < j< 0$ ( $\Theta_H>0$) case:\\
When $T_\infty > \Theta_H$, the number density will have a non-vanishing finite value outside the horizon. 
The number density around the horizon takes the form,
\be{n equil}
n \propto  \left(1-\frac{2M}{r}\right) ^{-\frac{c_v}{2}} .
\ee
When $T_\infty < \Theta_H$, the number density diverges at the surface of radius $r_0 (> 2M)$ given in Eq.~\eqref{r0:ig}. 
Inside this radius, the mild-heat flow approximation with $|Q| \ll 1$ may not hold.

\item 
$j=0$, (the thermal equilibrium, $\Theta_H=0$) case:\\
The number density takes an equilibrium form in Eq.~\eqref{n equil} which diverges at the horizon.

\item
$j> 0$ ($\Theta_H< 0$, heat out going) case:\\
The number density has a non-vanishing finite value outside the horizon.
Around the horizon, the number density takes the form of the equilibrium~\eqref{n equil}.

\end{itemize}

Putting these result to the first part of Eq.~\eqref{q:ig} with $\mu+\sigma=1+ c_v + m/\Theta$ , The heat behaves as
\be{heat:ig mild}
q = \frac{m+(1+c_v)\Theta}{4\pi r^2 \sqrt{1- 2M/r}} J_\infty
=\frac{\kappa M j}{r^2\sqrt{1-2M/r}}  \times \left\{
\begin{array}{cc}
\displaystyle   \frac{m}{1+c_v} + \Theta_H+(T_\infty-\Theta_H)
   \Big(1- \frac{2M}r \Big)^{-\frac{1+j}2}  ,   &   
j\neq -1, \vspace{.1cm}  \\
\displaystyle  
	 \frac{m}{1+c_v}+ T_\infty
   +  \frac{m}{2(1+c_v)} \log \left(1- \frac{2M}{r}\right)  , & j=-1 .    \\   
\end{array}
\right. 
\ee

\begin{itemize}
\item 
Asymptotically, the heat approaches $q \to q_\infty (r) =\left[m + (1+c_v)T_\infty\right]J_\infty/4\pi r^2 $.
\item
At $r=r_0$, the heat becomes
$
q(r_0) = \frac{mJ_\infty}{4\pi r_0^2\sqrt{1-2M/r_0}} ,
$
which flow direction is the same as that at the asymptotic region.
Because $q$ behaves monotonically, the flow direction does not change in the region where the mild-heat flow approximation is valid.
\end{itemize}

Now, we check the applicability of the low-boost approximation. From Eq.~\eqref{n mild},
\bea
|\epsilon| \approx 
 \frac{|J_\infty|}{4\pi n_0 r^2}
\left(1- \frac{2M}{r} \right)^{\frac{c_v-1}2 }  \times
\left\{
\begin{array}{cc}
\displaystyle  
	\left[1+ \frac{\Theta_H}{T_\infty} \left(1-\frac{2M}{r}\right)^{\frac{j+1}2 }
		-\frac{\Theta_H}{T_\infty} \right]^{-\frac{1+c_v}{j} }  ,   &   
j\neq -1, 0, \vspace{.1cm}  \\
\displaystyle 
	\left[ 1+\frac{m}{2 T_\infty (1+c_v)} 
	\log \left(1- \frac{2M}{r} \right)\right]^{(1+c_v)} ,  & j=-1 ,
	\vspace{.1cm}   \\ 
\displaystyle 
\exp \left[ \frac{m}{T_\infty} \left(1-\frac{2M}{r}\right)^{1/2} - \frac{m}{T_\infty}\right] ,  & \displaystyle 
j=0 .    \\   
\end{array}
\right.
\label{check low boost}
\eea
When $j \geq 0$ or $-1< j <0$ with $T_\infty > \Theta_H$, there is no position satisfying $n=0$ and  the near-horizon value of $\epsilon$ is proportional to 
$$
\epsilon \propto \left(1-\frac{2M}r\right)^{\frac{c_v-1}{2}} .
$$ 
This implies that the low-boost approximation is valid around the horizon because $c_v > 1$ for ordinary matter.
On the other hand, when $-1 < j < 0$ with $T_\infty < \Theta_H$ or $j \leq -1$, there exists a spherical surface $r=r_0 $ outside the horizon, where $\epsilon \to \infty$ because $n \to 0$.
Therefore, in this case, the low-boost approximation should be applied to the region $r > r_0$.

Now, we examine the viability of the mild-heat flow approximation when the low-boost approximation holds close to the event horizon, i.e., $j \geq 0$ or $-1 < j < 0$ with $T_\infty > \Theta_H$. 
For this purpose, we check the value of $Q$:
\bea \label{Q2 << 1}
|Q| 
&=&  \frac{ |\vartheta_0 J_\infty|}{4\pi n_0 \gamma c_v  T_\infty^2}  
	\frac{m+(c_v+1)\Theta_H
		+ (c_v+1)(T_\infty-\Theta_H)\left(1- \frac{2M}r\right)^{-\frac{j+1}{2} } }
		{r^2 \left(1- \frac{2M}{r} \right)^{-1-j +\frac{1-c_v}2 }  
			\left[1+ \frac{\Theta_H}{T_\infty} 
				\left(1-\frac{2M}{r}\right)^{\frac{j+1}{2} }
				-\frac{\Theta_H}{T_\infty} 
			\right]^{2+\frac{1+c_v}{j}} 
	 } \ll 1 
. 
\eea
Because we are considering region of parameters satisfying the low-boost approximation, the denominator does not vanish. 
Therefore, we can analyze the value around the horizon, which becomes $|Q| \propto (1- 2M/r)^{(c_v+j)/2}$.
Therefore, $|Q| \ll 1$ holds for all cases with $c_v > 1$ in the parameter region satisfying $j > -1$.
The mild-heat flow approximation is valid for all space regions outside the event horizon when $j \geq 0$ or $-1 < j < 0$ with $T_\infty > \Theta_H$.

\section{Analysis of the Ideal gas system with the low-boost approximation } \label{sec:V}
 
In this section, we examine the equation of motion~\eqref{eom1} which uses the low-boost approximation only.
Therefore, we do not assume $Q \ll 1$.
To simplify the discussion, we examine the massless particle case only.
Equation~\eqref{eom1} becomes, by using $q/\eta\Theta \approx 1+ c_v$, $\eta \approx n \epsilon$,  $N' = [2r(r/2M-1)]^{-1}$, and $N' +\eta'/\eta = -2/r$ from Eq.~\eqref{eta r}, 
\bea
\frac{r n'}{n} & \approx & 
	-\frac{2c_v F}{ c_v F-1} 
   +\frac{c_v }{c_v F-1}  \frac{1}{2(r/2M-1)},  \label{eom11} \\
\frac{r\Theta'}{\Theta} & \approx & 
 -\frac{1+j }{2 (r/2M-1)  }
-\frac{\gamma c_v}{1+ c_v }\frac{Q\vartheta}{\vartheta_0}
	\frac{B+1}{c_v F-1} 
		\left( \frac{c_v}{2(r/2M-1)}  -2 \right) .
\nn
\eea
Here, $j = \frac{J_\infty (1+c_v)}{4\pi \kappa M }$ as in Eq.~\eqref{aT}.
The two equations are a set of coupled non-linear differential equations for $n$ and $\Theta$. 

To analyze the equation, we combine the two equations to find a differential equation for one quantity $Q$.

From the definition of $Q$ in Eq.~\eqref{Q:def} and the first equality in Eq.~\eqref{q:ig},  we get,  $Q \propto q/n\Theta^2 \propto (\mu+\sigma)/ (r^2 e^N n \Theta)=(1+c_v)/ (r^2 e^N n \Theta)$ and $\frac{rQ'}{Q} = -rN'- 2 -rn'/n- r\Theta'/\Theta $.
By summing  the two equations in Eq.~\eqref{eom11}, 
\bea
y(y-1) \frac{d}{dy} \log Q  &=& \frac{j }{2 }
+\frac{2(y-1- c_v/4)}{ c_v F-1} \left[1
-\frac{2\gamma c_v}{1+ c_v }\frac{Q\vartheta}{\vartheta_0} (B+1)
		 \right],
\eea
where $y= r/2M$. 
Note that this is just a first order differential equation for $Q$ only, which can be solved easily at least numerically.
Finally, once we get $Q$ from this equation, we can also get the temperature equation (or the number density equation) by simply integrating  
\bea \label{dT:r1}
y(y-1) \frac{d}{dy} \log \Theta  &=&  
 -\frac{1+j }{2}
+\frac{2\gamma c_v}{1+ c_v }\frac{Q\vartheta}{\vartheta_0}
	\frac{B+1}{c_v F-1} 
		\left( y-1-\frac{c_v}{4}  \right) .
\eea
When  $Q$ and $\vartheta$ are $O(\epsilon)$, this equation reproduces the results in  subSec.~\ref{sec:4B1}.

\vspace{.3cm}
To go further, we need an explicit model for the higher order corrections of the heat behaviors. 
Here, we use 
\be{Phi}
\Phi(\vartheta) =\frac{\gamma}{2} \log \left(1+ \frac{\vartheta^2}{\vartheta_0^2} \right).
\ee
This choice reproduces the expansion in Eq.~(12) in Ref.~\cite{Kim:2023wel} to the quadratic order in $\vartheta$ and $\gamma'=0$ for the third order.
Now, from Eq.~\eqref{q:ig1}, we have 
$$
\frac{\vartheta}{\vartheta_0} = \frac{1- \sqrt{1- 4Q^2} }{2 Q} = \frac{2Q}{1+\sqrt{1-4Q^2}} , 
$$
where we choose negative signature in front of the square root.
With this choice, $\vartheta$ varies from $-\vartheta_0$ at $Q=-1/2$ to $\vartheta_0$ at $Q=1/2$.
Now, 
$$
B = \frac{\Phi'}{\vartheta \Phi''} =  \frac{\vartheta^2+\vartheta_0^2}{\vartheta_0^2 - \vartheta^2} = \frac1{\sqrt{1- 4Q^2}}  , 
\qquad 
F =  \frac{\Phi'^2}{\Phi''} = \frac{\gamma \vartheta^2}{\vartheta_0^2-\vartheta^2} = \frac\gamma 2 \frac{1- \sqrt{1- 4Q^2}}{\sqrt{1- 4Q^2}} .
$$
The differential equation for $Q$ becomes
\bea
y \frac{d Q}{dy}  
&=& - \frac{4\gamma c_v}{(1+ c_v)  (2+ \gamma c_v )}\frac{Q}{y-1}\frac{ \left(y- 1- \frac{c_v}4\right)   f( S)
   }{ S-\frac{\gamma c_v} {2+ \gamma c_v}   }, 
   \label{Q':y}
\eea
where $S \equiv \sqrt{1- 4Q^2} = B^{-1}$,
$$
f(S) = (S-S_+) (S-S_-)=  S^2 
 +\frac{1+ c_v } {\gamma c_v} \left(1  - \frac{2+ \gamma c_v }{8} \frac{j}{Y}\right)  S
	- \left(1- \frac{1+ c_v}{8}\frac{ j}{Y}  \right) ,
$$
and  $Y \equiv y- 1-c_v/4$. 
Here, $S_\pm $ are the two solutions of $f(S)=0$ with $S_+ < S_-$.
We name the corresponding values of $Q$ as $Q_\pm$ with $Q_+> Q_-$.
Note that $Q'=0$ at the points satisfying $Q=0$ and  $S= S_\pm$ satisfying $f(S) =0$.
Note also that the derivative $|Q'| \to \infty$ at the point $S = S_c$ with $S_c \equiv \gamma c_v/(2+ \gamma c_v)$, where $0  \leq S_c < 1$. 
This implies that as $Q$ increases from $0$ toward $1$, there exists a point satisfying $|Q'| \to \infty$, 
$$
Q_c = \pm\frac12 \sqrt{1- \left(\frac{\gamma c_v}{2+ \gamma c_v}\right)^2 }.
$$
Note that, at $Q=0$ ($S=1$), $Q' =0$ and 
$$
f(1) = \frac{1+c_v}{\gamma c_v} \left(1- \frac{j}{4Y}\right), \qquad
f'(1) = \frac{1+c_v + 2\gamma c_v}{\gamma c_v} 
	- \frac{(1+c_v)(2+\gamma c_v)}{8 \gamma c_v} \frac{j}{Y}.
$$
At $S=S_c$, 
$$
f(S_c) = \left(\frac{\gamma c_v}{2+ \gamma c_v } \right)^2 +\frac{1+c_v}{2+ \gamma c_v}-1, \qquad
f'(S_c) = \frac{2\gamma c_v}{2+ \gamma c_v }+\frac{1+ c_v } {\gamma c_v}-\frac{(1+ c_v)(2+ \gamma c_v ) } {8\gamma c_v}  \frac{j}{Y} .
$$
Interestingly, $f(S_c)$ is independent of $y$.
This presents an interesting possibility: If $\gamma$ is chosen to satisfy $f(S_c)=0$, one can smoothly go over the $Q=Q_c$ without experiencing the divergence of $Q'$.  

Let us consider the equation in the near-horizon limit $y \to 1$.  
If $Q(1) \neq 0, Q_\pm$ at $y=1$, $Q' \to \infty$ because of the $1/(y-1)$ factor in Eq.~\eqref{Q':y}.
Therefore, a regular solution at $y=1$ will be given only when we require $Q=0$ or $S = S_\pm$ there.
We first consider the case $Q(1) =0$ case. 
Taking the $y\to 1$ limit and using $f(1)_{y=1} = \frac{1+c_v}{\gamma c_v} \left(1+ \frac{j}{c_v}\right) $, the differential equation~\eqref{Q':y} gives
\bea
1
&=& \frac{1}{2}  \left(c_v+ j\right), 
   \label{reg cond}
\eea
Second, we consider $S(1)= S_+$ case.
 Equation~\eqref{Q':y} gives a constraint for $\gamma$, $j$ and $c_v$:
\bea
 1
&=& - \frac{\gamma c_v^2}{(1+ c_v)  (2+ \gamma c_v )}
	\frac{ (S_+ -S_-) (1-S_+^2)}
		{ S_+(S_+-\frac{\gamma c_v} {2+ \gamma c_v}  ) } .
   \label{Q':y2}
\eea

\vspace{.3cm}
Now, we analyze the differential equation~\eqref{Q':y} in the region outside the horizon, $y \geq 1$ i.e., $Y> -c_v/4$.
The gradient $Q'=0$ at the points satisfying $Q=0, ~Q_\pm$. 
Those points satisfying the conditions always belong to the solution curve. 
Therefore, it is important to examine the behavior of $Q$ around the points. 
Let us examine the behavior of $Q$ around $Q=0$ after setting $Q =  \delta Q \ll 1$:
\be{Q':y 1}
\frac1{\delta Q}\frac{d \delta Q}{dy}  
= -\frac{2}y + \frac{c_v+j}{2y(y-1)}
= -\frac{4(y-1) - c_v -j }{2y(y-1)} 
   .
\ee
Therefore, for the region $y \lessgtr 1+ (c_v+j)/4$,  $Q$ decreases/increases with $y$.
Integrating, the equation gives
\be{Q:r}
Q =\delta Q = Q_0 \left(\frac{2M}{r}\right)^{2} \left (1-\frac{2M}r \right)^{\frac{c_v +j}{2}} .
\ee
As seen here, when $c_v +j=2$, $Q$ and $Q'$ are regular at the horizon. 
When $0 \leq c_v +j< 2$, $Q$ is regular but $Q' \to \infty$ at the horizon. 
When $c_v+j > 2$, both $Q$ and $Q'$ vanishes at the horizon.
For $r \gg 2M$, $Q \ll 1$ because of the $(2M/r)^2$ factor. 
Therefore, only at the intermediate regions, $|Q|$ could be large. 
Especially, when $j \leq -c_v$, $Q$ decreases for all regions outside the horizon.  
This outcome strongly suggests that $Q= 0$ acts as an attractor with $r$.
The equation is derived based on the condition that $Q \sim 0$. 
Therefore, when the function $Q$ deviates from $0$ highly, the reliability of the results is questionable. 

We next examine the behavior of $Q$ around $Q_\pm$ after setting $Q= Q_\pm + \delta Q$. 
Then,
\be{B'3}
 \frac{d \delta Q}{dy}
= \mp \left[\frac{4 S_c}{1+ c_v}\frac{1 }{y(y-1)}\frac{  (S_-- S_+)   Q_\pm^2
   }{S_\pm( S_\pm -S_c )  } \right]
   \left( y-1- \frac{c_v+ j}4 \right)	\delta Q .
\ee
Note that the term in the squared bracked is non-negative. 
Therefore, $|\delta Q|$ will decreases/increases when $y \gtrless 1+ (c_v+ j)/4$.

\vspace{.3cm}
We next examine the behavior of the temperature.
The differential equation~\eqref{dT:r1} for temperature becomes
\bea
 \frac{d}{dy} \log \Theta  &=&  
 -\frac{1 }{2y(y-1)} \left[1+j
	+\frac{4 S_c}{1+ c_v}
	   \frac{4Q^2(y-1-\frac{c_v}{4}) }
		{\sqrt{1- 4Q^2}- S_c  } 
		 \right],
\eea
where $S = \sqrt{1-4Q^2} \geq S_c$.

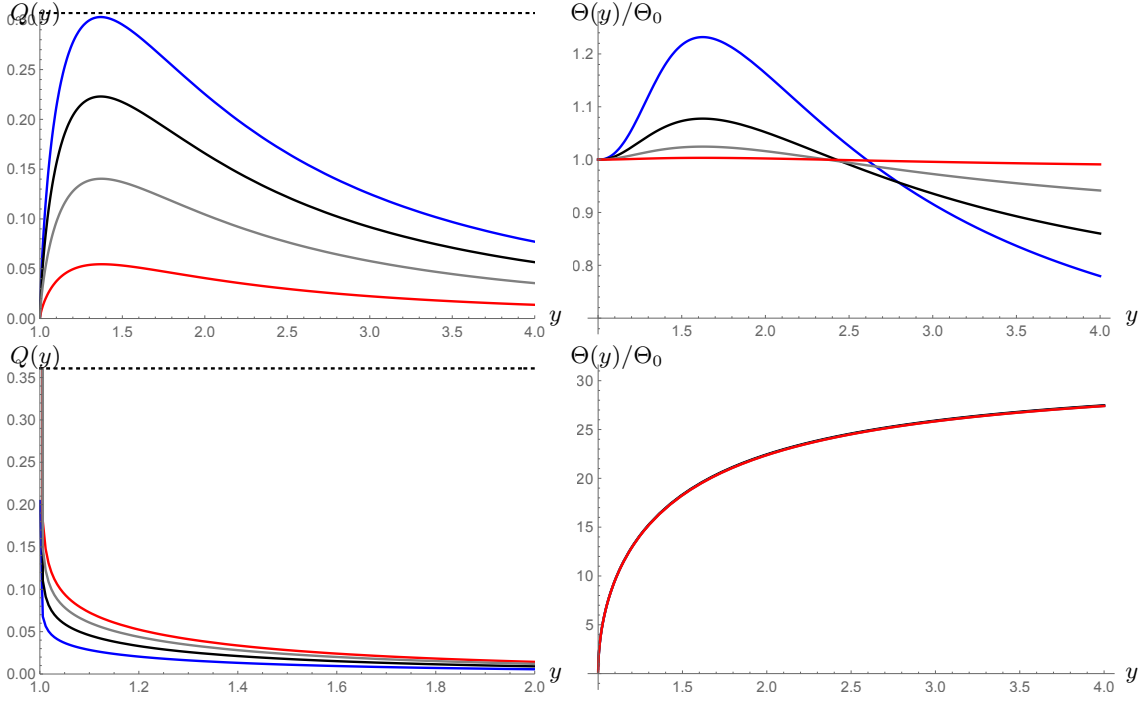
\begin{figure}[tbh]
\begin{center}
\begin{tabular}{cc}
\includegraphics[width=.4\linewidth,origin=tl]{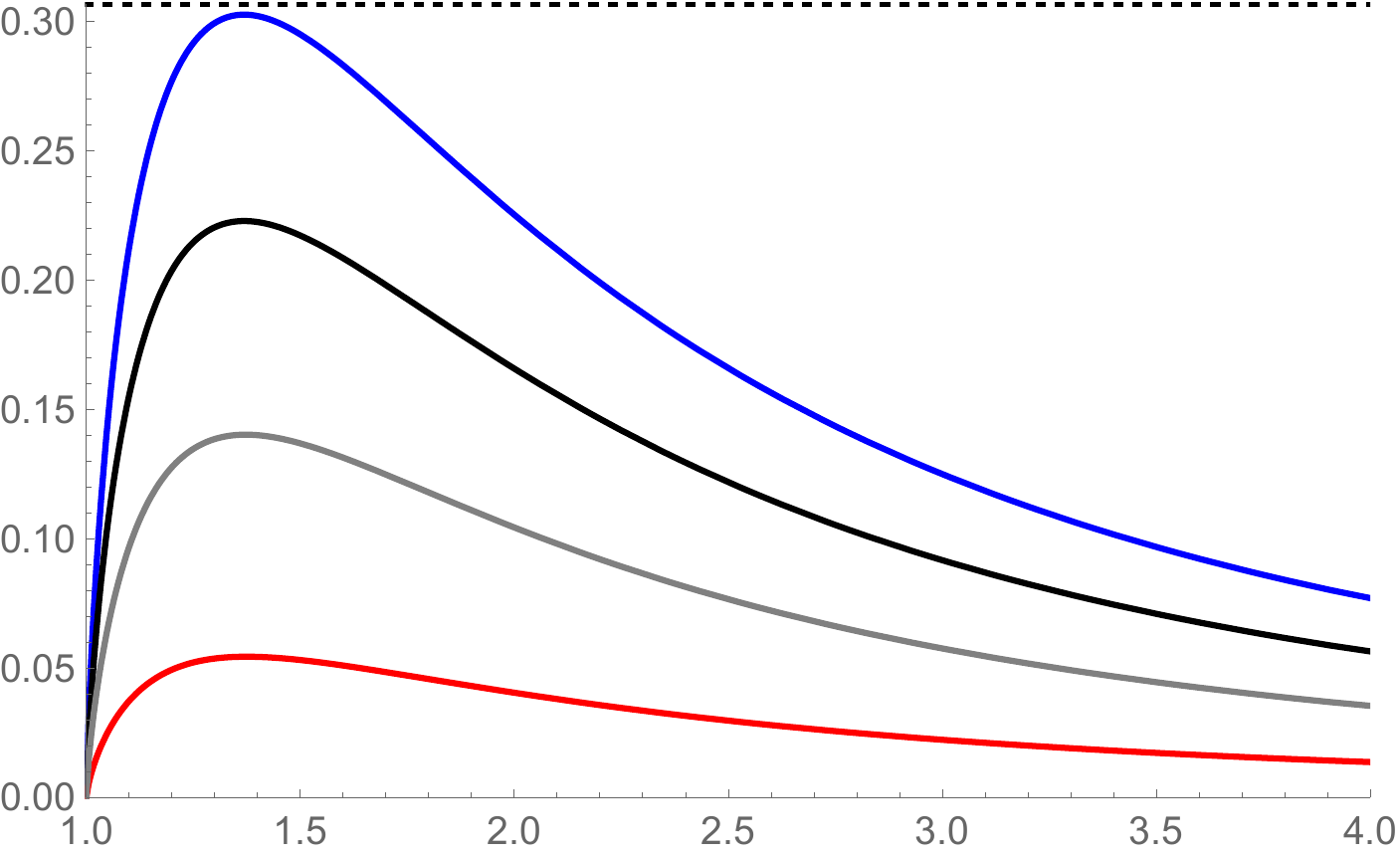} 
&
~~\includegraphics[width=.4\linewidth,origin=tl]{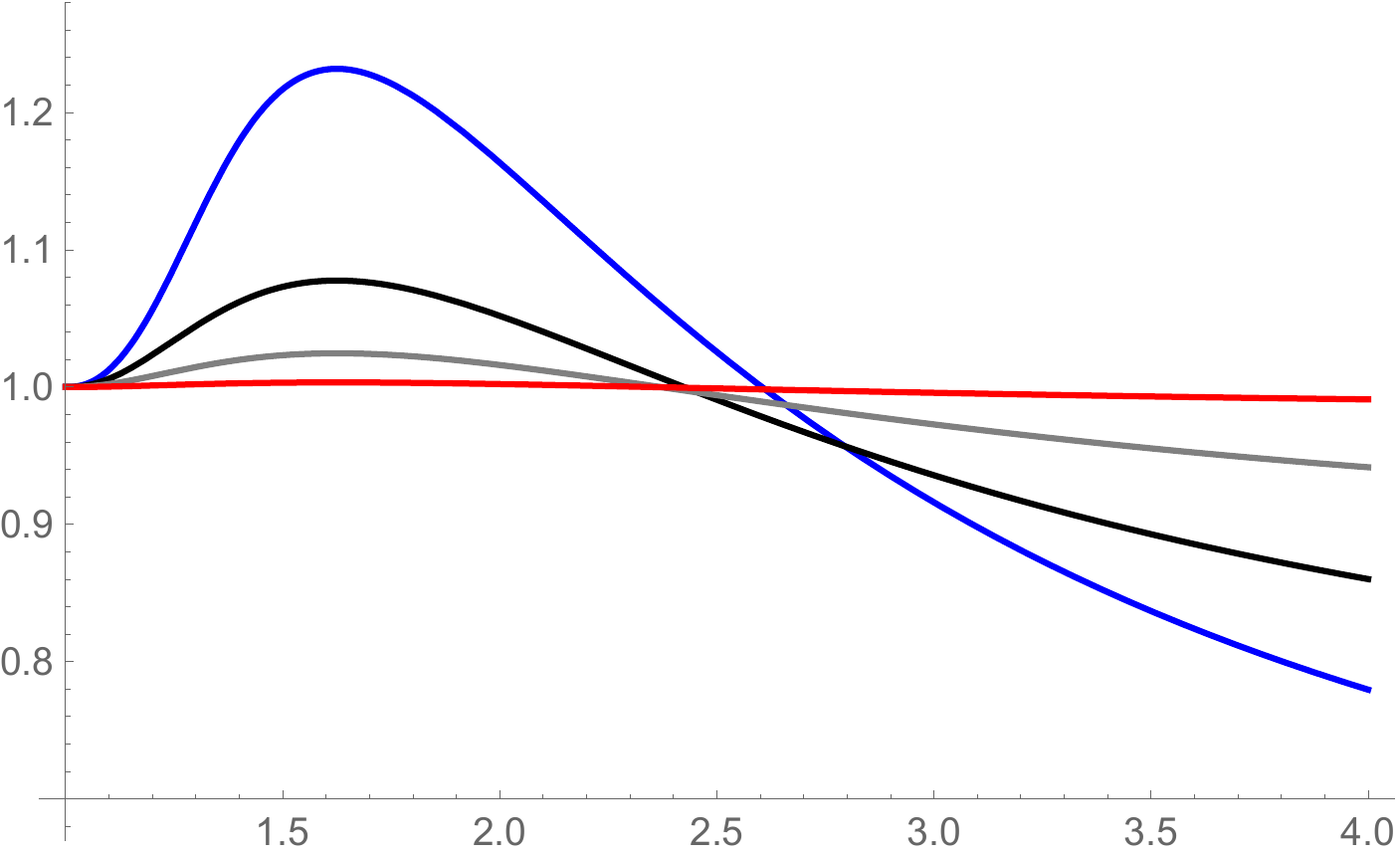} 
\vspace{.2cm}\\

\includegraphics[width=.4\linewidth,origin=tl]{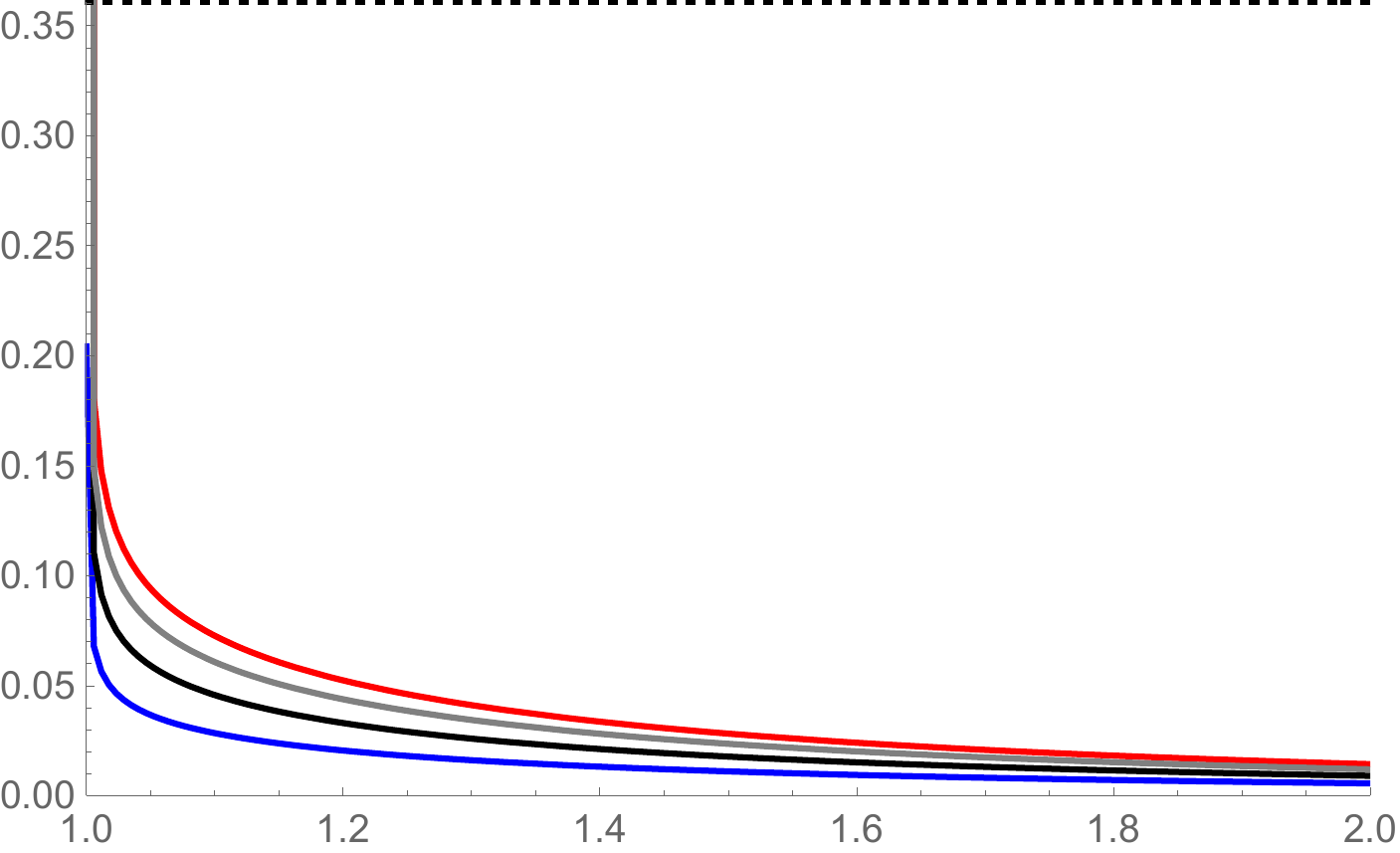} 
&
~~\includegraphics[width=.4\linewidth,origin=tl]{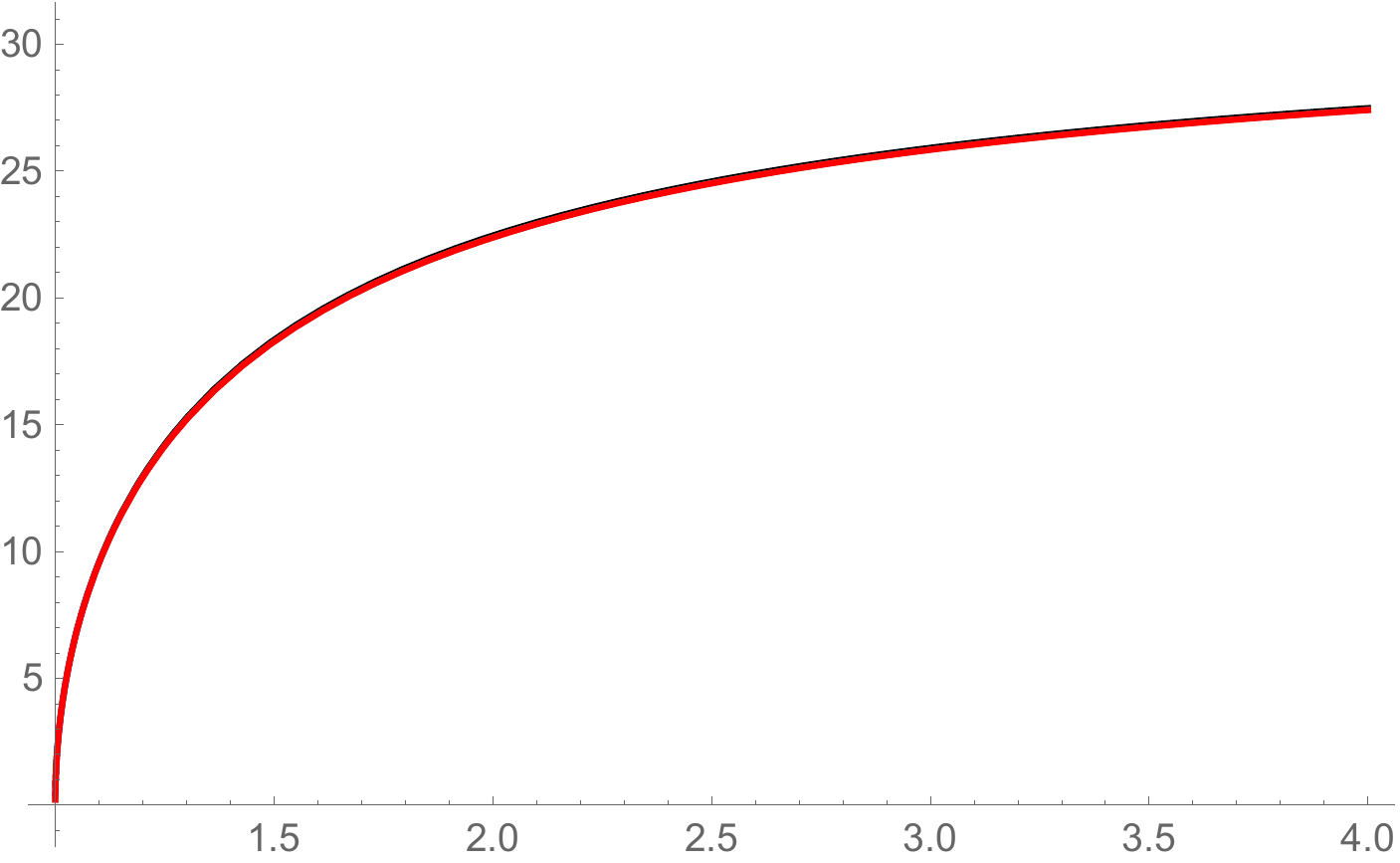} \\
\end{tabular}
\put (0,-119){$y$}
\put (0,16){$y$}
\put (-425,130){ $Q(y)$}
\put (-425,0){ $Q(y)$}
\put(-218, -119){$y$}
\put(-218, 16){$y$}
\put(-210,130){$\Theta(y)/\Theta_0$}
\put(-210,0){$\Theta(y)/\Theta_0$}
\end{center}
\caption{Numerical solution for $Q(y)$ (left) and $\Theta(y)/\Theta_0$ (right) for various initial conditions with $j=-1, ~c_v = 5/2$ (top) and $j=-2, ~c_v = 3/2$ (bottom).
 Here, we take $\gamma =1.5$ for both cases.
 Here, the black-dotted line denotes $|Q(y)| = |Q_c|$, the upper bound of the $|Q|$ value.
 We take the initial condition to be $Q'(1) = 29, 21, 13$ and $5$, (top) and $Q(1.001)/Q_c = 0.3 , 0.5, 0.7,$ and $0.9$ (bottom), respectively for blue, black, gray, and red curves. 
 The temperature curves in the bottom-right panel almost overlap with a minor difference (almost unnoticable in this figure).}
\label{fig:sol}
\end{figure}

When $Q$ is given by Eq.~\eqref{Q:r}, the temperature satisfies
\bea
 \frac{d}{dy} \log \Theta  =  
 -\frac{1 }{2y(y-1)} \left[1+j
	+\frac{4 S_c}{1+ c_v }
   \frac{4Q_0^2 y^{-4-c_v -j} 
   		\left[\left (y-1 \right)^{c_v +j+1} -\frac{c_v}{4} \left (y-1 \right)^{c_v +j}\right] }
	{\sqrt{1- 4Q_0^2 y^{-4-c_v -j} 
   		\left (y-1\right)^{c_v +j} }- S_c  } 
		 \right].
\eea
Integrating
\bea
\log \frac{\Theta}{\Theta_0} = - \frac{1+j}{2} \log (1-y^{-1}) 
-\frac{8 S_c Q_0^2}{1+ c_v }
\int^y dy' \frac{ {y'}^{-5-c_v -j} 
   		\left (y'-1 \right)^{c_v +j-1}  }
	{\sqrt{1- 4Q_0^2 {y'}^{-4-c_v -j} 
   		\left (y'-1\right)^{c_v +j} }- S_c  } 
		\left(y'-1-\frac{c_v}{4} \right) 
\eea
Because Eq.~\eqref{Q:r} naturally assumes $Q_0 \ll 1$, we may ignore the $Q_0^2$ term inside the root in the denominator of the integrand. 
Then, the local temperature becomes 
\bea
\Theta &\approx & T_\infty
(1-y^{-1}) ^{- \frac{1+j}{2} } 
\exp \left\{ \frac{4 \gamma c_v^2 Q_0^2   
	 }{1+c_v} \frac{(c_v+j-2)!}{y^{c_v+j+4}} 
	\right. \nn \\
&&	 \left. \times \left[(c_v+j-1)c_v\sum_{k=0}^4 \frac{3!}{(4-k)!} 
			\frac{(y-1)^{c_v+j+k} }{(c_v+j+k)!}
	 -\sum_{k=0}^5\frac{5! (y - 1)^{c_v+j+k-1} }{(5-k)!(c_v+j-1+k)!}\right] \right\} .
\eea
This result is consistent with Eq.~\eqref{massless} for $Q_0 \ll 1$.
In the large $y$, the exponent is governed by a term proportional to $(1-y^{-1})^{c_v +j+4} $. 
Therefore, $T_\infty$ becomes the asymptotic temperature. 
In the near horizon area with $y\sim1 $, the exponent takes the form, 
$$
- \frac{4\gamma c_v^2 Q_0^2}{1+c_v} \frac{(y-1)^{c_v+j-1} -1}{c_v + j-1} + O(y-1)^{c_v+j}
$$
When $c_v+j > 1$, this term presents a finite contribution to the temperature. 
When $j <-1$ additionally, the temperature vanishes at the horizon as shown in the bottom-right figure of Fig.~\ref{fig:sol}.
When $j=-1$ as shown in the top-right figure of Fig.~\ref{fig:sol}, the temperature has a non-vanishing finite value at the horizon.
When $c_v + j=1$, the term provides a logarithmic term of the form, 
$$
- \frac{4\gamma c_v^2 Q_0^2}{1+c_v} \log (1-y^{-1})+ O(y-1)^{1}
$$
Then, the temperature becomes
\be{Tf2}
\Theta \approx \Theta_0 \left(1-y^{-1}\right)^{-(1-\frac{c_v}{2}+\frac{4\gamma c_v^2Q_0^2}{1+c_v}) } + \cdots
\ee
Therefore, the local temperature vanishes/diverges at the horizon when $(1-\frac{c_v}{2}+\frac{4\gamma c_v^2Q_0^2}{1+c_v}) \lessgtr 0$.
Interestingly, the behavior of the local temperature depends on $Q_0$, which is related the strength of heat.

\section{Summary and Discussions} \label{sec:VI}
In this study, we examined the behavior of a steady system around a black hole event horizon compared to that of a thermal equilibrium system in the context of a spherically symmetric system governed by general relativity. 
We have focused on the steady state with radial heat conduction and analyzed the implications of heat flow on the system's thermodynamic properties.

The equations for radial heat flow in a steady state were previously investigated~\cite{Kim:2023lta}.
We reformulated those them in terms of key thermodynamic quantities such as the number density ($n$), the local temperature ($\Theta$), and the heat ($q$).
The system is modeled as a heat-flowing ideal gas, emphasizing deviations from thermal equilibrium and incorporating the explicit heat dependence of energy density as in Ref.~\cite{Kim:2023wel}.

Two approximations, namely the low-boost and the mild-heat flow approximations, were developed for analytic analysis. 
Exact solutions were derived based on the more stringent mild heat-flow approximation, particularly in a background Schwarzschild geometry. 
Notably, in the limit of vanishing mass for ideal gas particles, the behaviors of local temperature, number density, and heat simplify.

We found that when the heat flows in with the speed satisfying
$$
j \equiv \frac{(1+c_v)J_\infty}{4\pi \kappa M} \leq -1,
$$
the temperature at the horizon takes a finite value.
Here, $J_\infty$, $\kappa$, $M$, and $c_v$ denote the sum of the heat intensity over an asymptotic, spherical surface, the heat conductivity, the mass of the black hole, and the specific heat at constant volume, respectively.
When the inequality holds, the local temperature at the horizon vanishes. 
With an appropriate rate of heat inflow saturating the equality, the local temperature becomes finite and non-vanishing at the event horizon.

Furthermore, We explore the effects of non-linear behaviors of heat, finding an additional possibility for the horizon temperature to be finite. 
The study suggests that controlled heat flow alone, without considering the back-reaction of matter, can result in a finite local temperature at the horizon. 
The inclusion of matter's back-reaction to the black hole geometry remains an open and intriguing question.
This result prompts an optimistic expectation that considering this back-reaction appropriately may result in a finite local temperature for the system.

An interesting question arises when we apply the present analysis to the case of quantum mechanical Hawking radiation. 
It is usually regarded that the Hawking radiation originates from the event horizon to infinity. 
If this is right, the redshifted temperature $\Theta(r)/\sqrt{-g_{tt}(r)}$ must be higher for smaller $r$, making the temperature singularity at the horizon severe. 
A potential solution to this issue is that the starting position of Hawking radiation becomes indistinct due to the comparable typical wavelength of the radiation to the radius of the black hole.
Therefore, the local thermal equilibrium condition cannot be applied close the horizon.

In summary, the study provides insights into the thermodynamic behavior of a steady system around a black hole, challenging conventional expectations and paving the way for further exploration, especially concerning the interplay between matter and the black hole geometry.

%
\section*{Acknowledgment}
This work was supported by the National Research Foundation of Korea grants funded by the Korea government RS-2023-00208047.
\appendix


\end{document}